\documentstyle[aps,prl,epsfig,psfig]{revtex}
\tightenlines

   \newcommand{\be}{\begin{equation}}
   \newcommand{\ee}{\end{equation}}
   \newcommand{\bea}{\begin{eqnarray}}
   \newcommand{\eea}{\end{eqnarray}}
   \newcommand{\upar}{\uparrow}
   \newcommand{\dn}{\downarrow}
   \newcommand{\ek}{\varepsilon_k}
   \newcommand{\ef}{\varepsilon_f}

\begin{document}
\draft
\widetext
\title{Supersymmetric Approach to Heavy-Fermion Systems}
\author{C.P\'epin and M.Lavagna}
\address{Commissariat \`a l'Energie Atomique, \\ D\'epartement de
  Recherche Fondamentale sur la Mati\`ere Condens\'ee/SPSMS, \\
17, rue des Martyrs,
  \\ 38054 Grenoble Cedex 9, France}

\maketitle \widetext
  \leftskip 10.8pt
  \rightskip 10.8pt
  \begin{abstract}

We present a new supersymmetric approach to the Kondo lattice model in order to describe simultaneously the quasiparticle excitations and the low-energy magnetic fluctuations in heavy-Fermion systems. This approach mixes the fermionic and the bosonic representation of the spin following the standard rules of superalgebra. Our results show the formation of a bosonic band within the hybridization gap reflecting the spin collective modes. The density of states at the Fermi level is strongly renormalized while the Fermi surface sum rule includes $n_{c}+1$ states. The dynamical susceptibility is made of a Fermi liquid superimposed on a localized magnetism contribution. 

\end{abstract}

\bigskip


\noindent {\it keywords}: supersymmetry, heavy-fermions, strongly-correlated electrons, Kondo lattice, magnetism.

\bigskip 

Heavy-Fermion systems are typical examples of systems with strong electron-electron correlation effects. They are at the origin of the large effective masses as observed by specific heat, susceptibility and de Haas van Alphen measurements. The formation of the heavy quasiparticles is believed to result of the Kondo effect in these spin-fermion systems. At the same time, a serie of magnetic properties are found superimposed on the Fermi-Liquid type excitations. Let us quote the evidence for two distinct contributions in the dynamic structure factor measured in CeCu$_6$ and CeRu$_2$Si$_2$ by Inelastic Neutron Scattering: a q-independent quasi-elastic component typical of the local Kondo effect, and a strongly q-dependent inelastic peak reflecting the magnetic correlations due to RKKY interactions. Also the onset of a metamagnetic transition under magnetic field in CeCu$_6$ and CeRu$_2$Si$_2$, as well as the development of small ordered moments in UPt$_3$ are characteristic of local magnetism. Quite obviously, the physics of heavy-Fermions is dominated by this duality between Kondo effect and localized moments.

Traditionally, the spin is described either in fermionic or bosonic representation. If the former representation used for instance in the $1/N$ expansion of the Anderson~\cite{millis} or Kondo~\cite{auerbach} lattice, appears to be well adapted for the description of the Kondo effect, it is also clear that the bosonic representation lends itself better to the study of local magnetism. This constitutes our motivation to introduce a new approach to the Kondo lattice which relies on an original supersymmetric representation of the impurity spin $1/2$ in which the different degrees of freedom are represented by fermionic as well as bosonic variables. The concept of supersymmetry originally stems from field theory. It has been applied in condensed matter physics on two occasions: first in the problem of localization in disordered metals~\cite{efetov}, and secondly in the study of the underscreened Kondo impurity( $S>1/2$)~\cite{gan}. In the latter case,only the spin channel, which is screened by the conduction electrons ($s=1/2$), is treated fermionically, while the residual degrees of freedom are represented bosonically.
\par 
We propose the following supersymmetric representation of the spin $S=1/2$. 
The two states constituting the basis may be written: $|1/2,1/2\rangle =
(x f^{\dagger}_{\upar}+y b^{\dagger}_{\upar})|0\rangle$ and $ |1/2,-1/2\rangle = (x f^{\dagger}_{\dn}+y b_{\dn}^{\dagger})|0\rangle $, where $f^\dagger_\sigma$ and $b^\dagger_\sigma$ are respectively fermionic and bosonic creation operators, $x$ and $y$ are parameters controlling the weight of both representations and $\left| 0 \right\rangle$ represents the vacuum of particles: $f_\sigma \left| 0 \right\rangle = b_\sigma \left| 0 \right\rangle =0$.
The spin operators are then given by $ S^{\alpha} =  
\sum_{\sigma\sigma'}{f^\dagger_{\sigma}\tau^{\alpha}_{\sigma\sigma'}f_{\sigma'}
+b_{\sigma}^{\dagger}\tau^{\alpha}_{\sigma\sigma'}b_{\sigma'}}
 = S^{\alpha}_{f}+S^{\alpha}_{b}$, where $\alpha = \left(+,-,z\right)$ and $\tau^\alpha$ are Pauli matrices. ${\bf S}_f$ and ${\bf S}_b$ are similar to Abrikosov pseudo-fermion and Schwinger boson representation of the spin respectively.

One can easily check that the new representation satisfies the standard rules of $SU(2)$ algebra: $\left| S,m\right\rangle$ are eigenvectors of $S^2$ and $S^z$ with eigenvalues $S \left[S+1\right)$ and $m$, $\left[S^+,S^- \right] = 2 S^z$ and $\left[S^z,S^{\pm}\right]= \pm S^{\pm}$ provided that the following local constraint is satisfied:
\be \sum_{\sigma}{f_{\sigma}^{\dagger}f_{\sigma}+b_{\sigma}^{\dagger}b_{\sigma}}=
n_{f}+n_{b}=1 \ ,
\ee which implies $ x^2+y^2 =1$. In order to eliminate the unphysical states as $\left( x f^\dagger_\sigma + y b^\dagger_{-\sigma} \right) \left| 0\right\rangle$, we need to introduce a second local constraint and take: 
$Q=\sum_{\sigma} \sigma (f_{\sigma}^{\dagger}b_{-\sigma}+b_{-\sigma}^{\dagger}f_{\sigma})=0$ 

The partition function of the Kondo lattice can be written as a path integral: 
\be \begin{array}{l}
Z=\int {\cal D}c_{i\sigma}{\cal D}f_{i\sigma}{\cal D}b_{i\sigma}
d \lambda_{i} d K_i
\exp  \Bigl[-
  \int_{0}^{\beta} d\tau ({\cal L}_0 + {\cal H} \nonumber \\
   +\sum_i \lambda_{i}(n_{f_{i}}+n_{b_{i}}-1) + \sum_i K_i Q_i \Bigr] \ , \\  \\
\begin{array}{ll} 
\mbox{with} & {\cal L}_0=\sum_{i\sigma}c_{i\sigma}^\dagger\partial_{\tau}c_{i\sigma}+f_{i\sigma}^{\dagger} \partial_{\tau}f_{i\sigma} +b_{i\sigma}^{\dagger}\partial_{\tau}b_{i\sigma} \\ \\
\mbox{and} &  
{\cal H}=\sum_{k\sigma}{\ek c_{k\sigma}^{\dagger}c_{k\sigma}}
+\sum_{i}{E_{0}(n_{f_{i}}+n_{b_{i}})} \\
&  + J\sum_{i}{({\bf S}_{f_{i}}+{\bf S}_{b_{i}}).{\bf s}_{i}} \\
&  -\mu\sum_{i}{(n_{c_{i}}+n_{f_{i}}+n_{b_{i}})} \ , 
         \end{array} \end{array} \ee  
where $J$($>0$) is the Kondo interaction, and two time-independent Lagrange multipliers $\lambda_{i}$ and $K_{i}$ are introduced to enforce the local constraints.
\par
Performing a Hubbard-Stratonovich transformation and neglecting the space and time dependence of the fields introduced $\sigma_i$, $\lambda_i$, $\eta_i$ in a self-consistent saddle-point approximation, we have:
\be
\label{sadpoint} \begin{array}{l}
Z=\int{d\eta d\eta^{*}{\cal C}\left(\sigma_0, \lambda_0, \eta ,\eta*\right)Z(\eta,\eta^{*})} \\ \\
Z(\eta,\eta^{*})=\int{\cal D}c_{i\sigma}{\cal D}f_{i\sigma}{\cal D}b_{i\sigma}  \exp\Bigl[- \int_0^\beta d \tau \left( {\cal L}_0 + {\cal H}' \right) \Bigr] \ , \\ \\
 \begin{array}{ll}
\mbox{with} & {\cal C}\left(\sigma_0, \lambda_0, \eta \eta* \right) = \exp \beta \left( \lambda_0 - \frac{\sigma_0^2+ \eta^* \eta}{J} \right) \ ,\\
& {\cal H}' = \sum_{k\sigma}{(
f^{\dagger}_{k\sigma}c^{\dagger}_{k\sigma}b^{\dagger}_{k\sigma})
H_{0}\left(
\begin{array}{c}
f_{k\sigma} \\
c_{k\sigma} \\
b_{k\sigma}
\end{array}
\right)}  \ . \end{array} \end{array} \ee

$$\begin{array}{ll}
H_{0}=\left(
\begin{array}{ccc}
\ef- \mu & \sigma_{0} & 0 \\
\sigma_{0} & \ek- \mu & \eta \\
0 & \eta^{*} & \ef- \mu
\end{array}
\right) \ , & \ef=E_{0}+\lambda_{0} \ . \end{array}  $$

\par
Note the presence of a Grassmannian coupling $\eta$ between $c_{i \sigma}$ and $b_{i \sigma}$, in addition to usual coupling $\sigma_0$ between $c_{i \sigma}$ and $f_{i \sigma}$ responsible for the Kondo effect. $H_0$ is a supersymmetric matrix of the type $\left( \begin{array}{cc} a & \sigma \\ \rho & b \end{array} \right)$ in which $ a$,$b$ ($\rho$,$\sigma$) are matrices consisting of commuting (anticommuting) variables. 
\begin{figure}
\centerline{\psfig{file=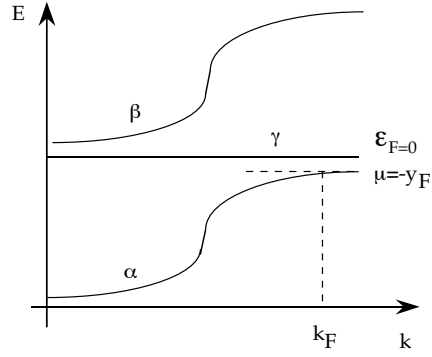,height=7cm,width=9cm}}
\caption{Sketch of energy versus wave number $k$ for the three bands $\alpha$, $\beta$, $\gamma$ resulting of the diagonalization of supersymmetric $H_0$. Note the presence of the bosonic band $\gamma$ separating the fermionic bands $\alpha$ and $\beta$. Note also the flatness of the lower band at $k=k_F$.}
\label{fig1}
\end{figure}

Next step is to diagonalize the supersymmetric $H_0$.
The resulting spectrum of eigenenergies is schematized in Figure ~\ref{fig1}. The new feature brought by the supersymmetric approach is the presence of a dispersionless bosonic band $\gamma$ within the hybridization gap separating the two fermionic bands $\alpha$ and $\beta$.
\par
In the scheme we propose, $\sigma_0$ and $\lambda_0$ are slow variables that we determine by solving the saddle-point equations, while $\eta$, $\eta^*$ are fast variables that are defined from a local and instantaneous approximation. Performing the functional integration of~(\ref{sadpoint}) over the fermion and boson fields yields~\cite{efetov} a superdeterminant ($SDet$) form written as follows

$$ \begin{array}{l}
Z(\eta, \eta^*) = SDet(\partial_\tau + H) \ , \\ \\
   \begin{array}{ll}
\mbox{where} &  {\displaystyle SDet(\partial_\tau + H) = \frac{Det(G^{-1} - \sigma D \rho ) }{ Det( D^{-1})} }\ , \end{array} \\
\begin{array}{lll}
G^{-1}= \partial_\tau + a  & \mbox{and} & D^{-1} = \partial_\tau + b \ .
 \end{array} \end{array} $$

Expanding to second order in $\eta$, $\eta^*$ allows us to define the propagator $G_{\eta \eta^*}( {\bf k}, i \omega_n) $ associated to the Grassmann variable $\eta$ and hence the closure relation for $x_0^2 = \left \langle \eta \eta^* \right\rangle$:
\be
\label{4}
{\displaystyle x_{0}^{2}=\frac{1}{\beta}\sum_{{\bf k},i\omega_{n}}G_{\eta \eta^*}( {\bf k}, i \omega_n)} \ ,
\ee 
$$ \begin{array}{ll}
\mbox{with} & {\displaystyle  G_{\eta \eta^*}( {\bf k}, i \omega_n) = \frac{J}{
\left[1- J\Pi_{cb}^{0}(\bf k,i\omega_{n})\right]} }\\
\mbox{and} & {\displaystyle \Pi_{cb}^0 = \frac{1}{\beta} \sum_{{\bf q}, i \omega_n} G_{cc} ({\bf k+q}, i \omega_n + i \omega_\nu) D( {\bf q}, i \omega_n) }\ .
\end{array}  $$

\begin{figure}
\centerline{\psfig{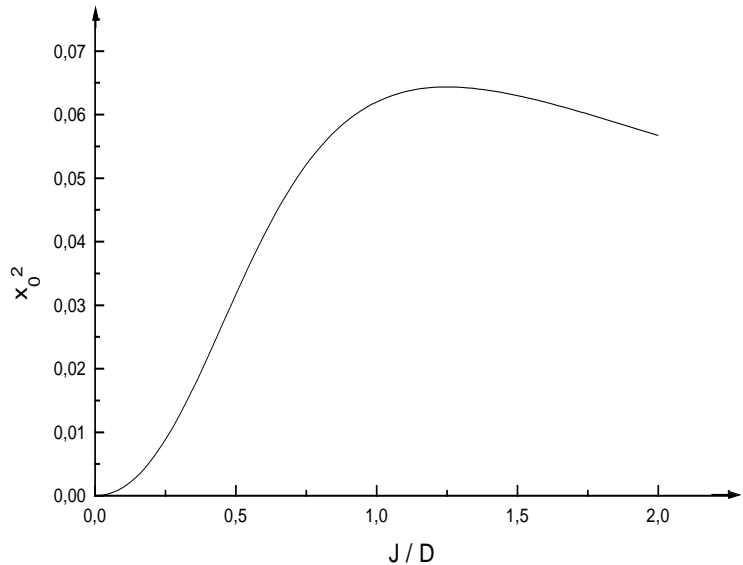}}
\caption{$J/D$-dependence of the renormalized Grassmannian coupling $x_0^2= \protect \left\langle \eta \eta^* \protect \right\rangle$, fixing the weight of fermion and boson statistics.}
\label{fig2}
\end{figure}

The resolution of the saddle-point equations, keeping the number of particles conserved, leads to:
\be \label{5} \begin{array}{c}
{\displaystyle -y_{F}=(\ef-\mu)=D \exp\left[-\frac{1}{2J\rho_{0}} \right] } \ , \\
{\displaystyle 1= \frac{2 \rho_0 ( \sigma_0^2 + x_0^2)}{- y_F} }\ ,  \\
{\displaystyle \mu = - \frac{(\sigma_0^2 + x_0^2)}{D} }\ ,   
\end{array} \ee where $\rho_0 = 1/2D$ is the bare density of states of conduction electrons. From this set of equations we find : $\ef =0$ and $\mu = y_F$. These equations constitute an extension of the usual mean-field results of
the 1/N expansion~\cite{millis,auerbach} to the supersymmetric case. A key role is played by the parameter $x_0^2$ determined by the closure equation~(\ref{4}) that controls the weight of bosonic over fermionic
statistics. We have studied the J-dependence of $x_0^2$ and reported the results in Figure ~\ref{fig2}. $x_0^2$
first increases when $J$ is reduced and then decreases reaching zero at $J=0$. 

At zero temperature, only the lowest band $\alpha$ is filled.  Let us then discuss the density of states $\rho(E_F)$  and the respective weight of $f$, $c$ and $b$ in the quasiparticle $\alpha$ formed at the Fermi level. A na\"{\i}ve argument might lead us to think that $\rho(E_F)$ could be depressed when gaining new bosonic degrees of freedom. In fact nothing of the sort happens: we found an enhancement of the density of states unchanged from the standard $1/N$ expansions: $ {\displaystyle \frac{\rho(E_F)}{\rho_0} = 1+ \frac{(\sigma_0^2 + x_0^2)}{y_F^2} \simeq 1+ \frac{D}{(-y_F)} \ll 1 }$. The weight of $f$,$c$,$b$ in the quasiparticle $\alpha$ at the Fermi level are  given by $u_k^2$,$v_k^2$ and $\rho_1 \rho_1^*$ respectively.  The bosonic weight behaves as $x_0^2$ as a funcion of $J$. Note that all of these results are valid as long as the bosons are not condensed.

Largely discussed in the litterature \cite{ueda} is the question concerning the Fermi surface sum rule: do the localized spins of the Kondo lattice contribute to the counting of states within the Fermi surface or do they not? Depending on the answer, one would expect large or small Fermi surfaces. The supersymmetric theory leads to a firm conclusion in favour of the former. It is easy to check that the number of states within the Fermi surface is just equal to $n_c+n_b+n_f$, i.e. $n_c+1$. The Luttinger rule includes both conduction and localized electrons, even if the latter described as spins in the Kondo hamiltonian are deprived of charge degrees of freedom.

Let us next consider the dynamic spin susceptibility. It consists of two contributions, one fermionic $\chi_{ff}$ and one bosonic $\chi_{bb}$. Writing them in the eigenbasis $\{ \alpha, \beta, \gamma \}$ and neglecting the interband term in the frequency range of interest, one gets: $\chi({\bf q}, \omega) = \chi_{\alpha \alpha}({\bf q}, \omega) + \chi_{\gamma \gamma}({\bf q}, \omega)$, with 
\be \begin{array}{l}
{\displaystyle \frac{\chi_{\alpha \alpha}({\bf q}, \omega)}{\omega} = \rho_0 \left(1+ \frac{\sigma_0^2}{y_F^2} \right) \left[ \frac{1}{\omega} + i \frac{a}{ q v_F^*} \right]} \ , \\
{\displaystyle \frac{\chi_{\gamma \gamma}({\bf q}, \omega)}{\omega} =\frac{1}{4 T \sinh^2(\frac{\beta T_K}{2} )} \left[ \frac{1}{\omega} + 2 i a \delta(\omega) \right] }\ , \end{array} \ee
$a= \pi/2$ for $ \ek \simeq k^2/2 m$ and $v_F^*$ is the Fermi renormalized velocity. 
The first contribution is of Lindhard-type. It is characteristic of fermionic quasiparticles. The second contribution is bosonic. It is zero at $T=0$ since the $\gamma$-band is empty. It increases as the temperature gets larger, with a peak of $\chi^{\prime \prime}(\omega)/\omega$ at $\omega = 0$.

In summary, we have shown that the supersymmetry proves to be a powerful tool 
already at mean-field level to account for both the Kondo effect and the localized magnetism present in the heavy-Fermion systems. We show how the weight of each of the bosonic and fermionic component is controled by a Grassmannian coupling. Following the standard rules of superalgebra, we find an increasing weight of bosonic component when the Kondo coupling is reduced before decreasing and reaching zero at $J=0$. The key results are the prediction of two contributions in the dynamical spin susceptibility, of Fermi liquid and localized magnetism-type respectively, the existence of an enhanced density of states at the Fermi level reflecting the heavy effective mass while the Fermi surface sum rule includes the total number of $n_c +1$ states.

\par

\end{document}